\documentstyle[12pt]{article}
\begin{document}
\vskip 2 cm
\begin{center}
\large{\bf
AN " INSTRUMENTALISM TO REALISM " HYPOTHESIS}
\vskip 3 cm
{\bf Afsar Abbas}\\
Institute of Physics, Bhubaneswar - 751005, India\\
{afsar@iopb.res.in}
\vskip 4 cm
{\bf ABSTRACT}
\end{center}
\vskip 2 cm

It is proposed here that all successful and complete 
theories always proceed through an intermediate
stage of instrumentalism to the final stage of realism.
Examples from history of science ( both classical and modern )
in support of this hypothesis are presented.

\newpage

Let us take instrumentalism as a representative 
antirealistic theory of science. 
It asserts that the scientific theories are merely instruments
whose value consists in their ability to successfully explain and
predict the experimental outcomes rather than in their fundamental
structure of reality (Ladyman (2002)). Realism on the other hand
posits fundamental structure and reality to objects and concepts
including mathematical ones.

Here I wish to propose the hypothesis that all successful theories  
necessarily go
through an intermediate "instrumental" stage to the final
"realistic" stage. So to say all good theories have to pass
through an antirealistic stage to the ultimate realistic stage.

The heliocentric theory of the planetary motions of Nicholaus
Copernicus was published posthumously in 1543. The book included a
preface by Andreas Oslander who was a friend of Copernicus and
who was instrumental in getting the book published. The preface was 
written by Oslander. In trying to explain the motivation,
fundamental basis and philosophical ideology of the heliocentric
theory of Copernicus, he said that the motion of earth need not be
taken literally. He said it was a convenient assumption to explain
the motion of planets. According to him it was just a mathematical
fiction which was apparently serving the useful purpose of
consistently explaining the physical reality of planetary motion.
In general, the historians of science 
have interpreted this to mean that it
was done so as not to offend the church too much. Hence as per
this view, it was a conscious attempt on the part of Oslander
to appease the church and the contemporary prejudice so as to 
make it more acceptable to them.
 
In the light of the facts which will emerge as we study some
other example of theories below, I would like to suggest that the
reason that Oslander wrote the above preface was perhaps because he
( and perhaps even Copernicus ) actually believed in the
"mathematical fiction" idea. 
There are reasons why this point of view may be taken seriously.
In fact Bertrand Russell ( Russell ( 1946 ), p. 512 ) says,
" It is uncertain how far Copernicus sanctioned this statement,
but the question is not very important, as he himself made similar
statements in the body of the book. The book is dedicated to the
Pope, and escaped official Catholic condemnation until the time
of Galileo."  He pointed out that Copernicus was an ecclesiastic of 
unimpeachable orthodoxy and that, " .. Copernicus .. protested
.. against the view that his theory contradicted the Bible."
As an honest Catholic and an upright mathematician/scientist
the only way that he could have held on to the above view was
by sincerely believing that 
his theory was actually a convenient "mathematical fiction"
say, to provide a better calender.

This point of view is further consolidated by the following facts.
Rupert and Mary Boas Hall ( Hall and Hall (1964) p. 140, 144 )
point out that it was common knowledge among Renaissance
astronomers that the Ptolemy's system was actually quite
complicated mathematically
and that it did not entirely agree with the 
Aristotelean universe either. Its tables contained errors and
sometimes its predictions were off by weeks. It also failed to
teach as to how to make corrections in its calender.
"This last problem, of so great concern to church, caused the
papacy itself to bless the search for new system".
Others besides Copernicus, though not as successfully, also tried
to formulate new systems to improve upon Ptolemy's system.
In fact, "Copernicanism was not suppressed ( one of the friends
who urged Copernicus to publish was a cardinal, the other a
catholic bishop )" ( Hall and Hall (1964) ).

Ever since the dawn of human
civilization mathematics has always been placed at the highest
intellectual level of human thought. Mathematics existing as
absolute entity in some putative Platonic world ascribes to this
point of view. 
As to how this Platonic world "dirties"
its hands by explaining the "mundane" physical reality has been a
puzzle for scientists. Even the mathematical physicist 
Eugene Wigner ( Wigner (1960) ) was compelled to state, 
"The miracle of the appropriateness of the language of 
mathematics for the foundation of the laws of physics is a 
wonderful gift which we neither understand nor deserve".

Thus "miracle" is nothing more than 
"the mathematical fiction" which
Oslander talks about. Such a way of treating a theory as a
"mathematical fiction" in trying to make a connection between the
Platonic world of 'real' mathematics with the fuzzy world of
physical reality in which some of the beautiful mathematics
finds application in. This is how they reconciled 
the dichotomy of pure and applied mathematics. 
This problem has persisted until now
and no wonder at the early stages (or even to very late
stages) application of a mathematical language which seems to
map and explain a physical reality could be considered a
"mathematical fiction". Hanging on to this concept of
"mathematical fiction" was perhaps one of the fundamental reason
why idealism/antirealism of Berkeley and others 
could sustain itself. 

Later Newton developed a more 
consistent theory of gravity which was
found to have a wide and accurate range of applicability and as
further observations were made, the "mathematical fiction" of
Copernicus started losing appeal. 
It started looking less of 
a fiction to others and appeared to become more "realistic". 
And today, the Copernican theory is considered to be as 
exact and as real as that any scientific theory can be. 
 
Another example of how a good scientific 
and successful theory would go from an
instrumental stage to the stage of realism is provided by Dalton's
atomic theory ( Shapere (1964) ). 
Dalton's  theory was never fully
accepted by his fellow scientists. For example Oswald and Davy and
others regarded Dalton's atomic theory with positive dislike and
were always looking for an appropriate substitute. At best they
were willing to accept "atom" 
as a convenient fiction which served the
purpose of explaining some physical reality. Often they would skip
the word "atom" altogether and use other contrived terms like 
"proportions" or "equivalents". In fact when a Royal Medal was
presented to Dalton ( Shapere (1964) ) the citation stated that it
was for his development of the theory of Definite Proportions
usually called the Atomic Theory of Chemistry. This clearly shows 
that in the initial stages the Atomic Theory was 
instrumentalistic in
as much as that was the way it was 
actually perceived by all the scientists. 
Today, clearly when scientists have actually "seen" atoms of 
the size of a few Angstrom, the atoms are as 
real as the pen in my hand. 

Another example of a physical theory undergoing instrumental
to realistic "transition" is that of Kekule's visualization of the
model for Benzene (C6 H6). He claims to have seen it in his dream.
He visualized it as a closed chain of six carbon atoms with six
hydrogen atoms hanging from 'strings' from the corresponding
carbon atoms. He or his contemporaries did not associate with
this picture any actual reality 
of existence in the three dimensional space.
For them it was just a convenient model - an instrument to
mock up the existence of Benzene and its properties. 
Later this structure became real when Van Laue demonstrated 
its actual physical structure in three dimensional space.

The next example is from modern particle physics / 
High Energy Physics.  Until 1930's
scientists were aware of the existence of electron, proton
and perhaps photon as the only "elementary particles" of which all
the other entities in nature were made up of. Thereafter neutron,
mesons and several other objects as elementary entities started
manifesting themselves in laboratories. So much so that by late
1950's and early 1960's there were close to about one hundred such 
entities which could be labelled as "elementary". There was
tremendous confusion in physics and new ideas were  cooked up to
explain this proliferation of particles. One such model was named 
"nuclear democracy" which in a way provided 
elementarity and equality to 
all these multitude of particles. 
However, the physicists were forced to abandon this model
as further empirical information seemed to contradict the
predictions of this model.

Protons and neutrons are massive particles whose electric charges 
are one and zero respectively ( in units of negative of electronic
charge ). To account for the existence of the large number of so
called "elementary particles", in the early 1960's, M. Gell-Mann
and G. Zweig independently proposed that these were actually
built up of more elementary entities (now) called quarks. Quarks
were of three different kind and named as: 
u (up- quark) d (down-quark) and s (strange-quark). 
These had electric charges 2/3, -1/3 and -1/3
respectively. Proton was made up of 3 quarks : u+u+d whereas
neutron was made up of 3 quarks : u+d+d. This description was
provided in the mathematical framework of the
group SU(3). These quarks were proposed to be the "elementary
particles" of which all the other strongly interacting particles
known then were made up of. However both Gell-Mann and Zweig 
did not posit any physical reality onto these quarks. As per
them - these quarks were just "mathematical entities" 
and SU(3) was a convenient mathematical "trick" 
cooked up to do the job. And indeed,
it did a good job of providing a comprehensive and consistent
description of reality which hitherto 
had appeared to be  quite bizarre. 
When Gell-Mann was later awarded Nobel prize in physics in 1969
for this work it was cited as 
"for his contributions and discoveries
concerning the classification of elementary particles and their
interaction".

During the Nobel Prize Ceremony Gell-Mann did present a lecture
entitled, " Symmetry and currents in particle physics". On Dec
11, 1969 at Stockholm. All Nobel Laureates are supposed to present
a written version of these lectures to be published. But for some
inexplicable reason Gell-Mann Nobel Lecture was not written up
for publication in the collection of lectures 
( Nobel Lecture (1963 - 1970) ). However it is 
reputed that he had actually referred to quarks as mere
"mathematical entities" in his lecture. Also a Presentation Talk
is given by a Nobel Committee Member before the award of the
Nobel Prize. This was done by Ivar Walter whose talk is present in
the Nobel Lecture Physics 1970, p 297. 
I quote Walter , "The
quarks are peculiar in particular because their charges are
fraction of the proton charge which according to all experience
up to now is the indivisible elementary charge. It has not yet been
possible to find individual quarks although they have been eagerly
looked for. Gell-Mann's idea is none the less of great heuristic
value". 

This conveys the idea of a "mathematical fiction" as clearly as
possible. For Gell-Mann and Zweig these quarks were mathematical
entities/fiction just to get the mathematical symmetries/group theory
right.  Zweig has said 
( Zweig (1965), p192 ), "In fact, it is likely
that the fundamental objects do not correspond to physical
particles; rather, the units may form a convenient set of symbols
that are helpful in expressing certain symmetries of the strong
interaction". 

Hence it is clear that the quarks in this model were for a long time
believed to be mere "mathematical entities" whose role was
basically to provide convenient representations of the
SU(3) group to describe the empirical facts in a consistent
manner. This was the "instrumental" stage of the development of this 
theory. Today the same quark model is on a very solid footing. 
In fact, the 1990 Nobel Prize in physics was 
awarded to J. I. Freedman,
H. W. Kendall and R. E. Taylor for basically experimentally 
demonstrating the physical reality of these quarks. So today
quarks are as real to a physicist as any particle can be.

If one studies any particular mature scientific theory carefully, I am
sure, one will discover the same pattern as above. 
The theory would finally become realistic only after going through the
initial or intermediate stage of being instrumentalistic. 
Hence I propose that this "Instrumentalism to Realism Hypothesis" 
may actually be a "law". However, sometimes this transition may 
be achieved after decades or even centuries. 

One should be warned though that there may be 
popular scientific theories today 
which are still in the instrumental 
stage. For such a theory the final realistic theory is yet to see
light of the day. As such that particular theory should be
considered tentative at best - as per the prediction of the 
"Instrumentalism to Realism" hypothesis presented here.

One important example of such a theory is quantum mechanics.
Quantum mechanics, in spite of unequivocal successes has
fundamental interpretational problems: non-locality, 
collapse of wave function and quantum jumps 
to name a few. As such many scientists, 
working in the field today, have come around to endorse the view
that quantum theory at present 
is only an "instrumental" theory. 
Let me quote Roger Penrose ( Penrose ( 2004 ) ),
"In this chapter, I shall put the case to the reader that there
are positive powerful reasons, over and above the negative ones
put forward in the preceding chapter, to believe that the laws of
present day quantum mechanics
are in need of a fundamental ( though presumably subtle )
change. These reasons come from within accepted physical
principles and from observed facts about the universe".

And indeed, as per my hypothesis presented here, 
this is inevitable. 
Note that the hypothesis presented here does not only have an
explanatory value, but it also has a predictive value as well.
And that can help in giving direction to science. This is 
in spite of the fact that
most of the working scientists feel that the discipline of 
philosophy of science
is completely irrelevant to their work.
The present hypothesis tells scientists not to 
become too pessimistic.
Many a working scientist in quantum mechanics appear to 
have done exactly this! 
This is due to a common feeling among scientists 
that the Copenhagen
interpretation of quantum mechanics ( the most popular one )
leaves them with hardly any other choice.
As per the hypothesis presented here, 
they should keep looking as the final
stage of any good theory is the realistic stage and that is yet
to be achieved in quantum mechanics.

\newpage

\vskip 2 cm
{\bf REFERENCES}
\vskip 1 cm

Hall, A. R, and Hall, Marie Boas (1964), 
" A brief history of science ",Signet Science Library Books, 
The New American Library of World Literature, New York

Ladyman, J. (2002), " Understanding philosophy of science ",
Routledge, London

Nobel Lectures - Physics (1963 - 1970), Elsevier Publishing
Company, New York 

Penrose, R. (2004), " The road to reality: a complete guide to 
the laws of the universe ", Jonathan Cape, London

Russell, B. (1946), " History of western philosophy ", 
Routledge, London

Shepere, D. (1964), " The structure of scientific revolutions ",
Philosophical Review, Vol. 73, 383 - 94

Wigner, E. (1960), " The unreasonable effectiveness of 
mathematics in the natural sciences ", Communications in Pure 
and Applied Mathematics, Vol 13, No. 1

Zweig G. (1965), " Fractionally Charged particles and SU(6) ", 
in " Symmetry in Elementary Particle Physics ", Ed. A. Zichichi,
Holt Academic Press, New York

\end{document}